\documentclass[intlimits,twoside,a4paper]{article}

\usepackage{amsmath,amssymb}
\usepackage{graphicx}
\usepackage{wrapfig}

\usepackage[T2A]{fontenc}
\usepackage[cp1251]{inputenc}
\usepackage[eqsecnum]{cmpj2}

\issue{2012}{15}{2}{23004}

\doinumber{10.5488/CMP.15.23004}

\title[Extrapolation of virial coefficients of HS]%
{On extrapolation of virial coefficients of hard spheres%
\thanks{Dedicated to Professor Orest A. Pizio on the occasion of his 60th birthday}}
\author[M.~On\v c\'ak \textsl{et al.}]{%
  M.~On\v c\'ak\refaddr{A1}, A.~Malijevsk\'y\refaddr{A1,A2}\thanks{E-mail: anatol.malijevsky@vscht.cz}\,, J.~Kolafa\refaddr{A1}, S.~Lab\'\i k\refaddr{A1}}

\addresses{
\addr{A1}{Institute of Chemical Technology, Prague, Department of Physical Chemistry, 166~28 Praha~6, Czech Republic}
\addr{A2}{Department of Physics, Faculty of Science, University of Ostrava, 701 03 Ostrava 1, Czech Republic}
}

\date{Received January 31, 2012, in final form April 18, 2012}
\authorcopyright{M.~On\v c\'ak, A.~Malijevsk\'y, J.~Kolafa, S.~Lab\'\i k, 2012}
\begin{document}

\maketitle

\begin{abstract}
 Several methods of extrapolating the virial coefficients, including those
  proposed in this work, are
 discussed. The methods are demonstrated on predicting higher virial
 coefficients of one-component hard spheres. Estimated values
  of the eleventh to fifteenth virial coefficients are suggested.
It has been speculated that the virial coefficients, $B_n$, beyond
$B_{14}$ may decrease with increasing $n$, and
   may reach negative values at large $n$. The extrapolation techniques may be utilized in other fields of science
   where the art of extrapolation plays a role.
\pacs 02.30.-f, 51.30.+i
\end{abstract}

\section{Introduction}
Virial coefficients play a key role in both molecular and
phenomenological theory of fluids at low and medium densities.
They are the coefficients in the density expansion of the equation
of state expressed via the compressibility factor
\begin{equation}
  Z=\frac{\beta P}{\rho}=1+\sum_{i=2}^\infty B_i\rho^{i-1}\label{39}\,,
\end{equation}
where $\beta=1/(k_{\rm B}T)$ is the inverse temperature, $P$ is
the pressure, and $\rho$ is the number density. Virial
coefficients $B_i$ are defined by exact formulae in terms of
integrals whose integrands are products of Mayer functions
\cite{hill}.

In this paper we concentrate on the simplest non-trivial model of
fluid -- the system of hard spheres. For hard spheres, i.e., hard body one-component fluids, the virial coefficients
are numbers (they do not depend on temperature). For the system
considered, the terms up to $B_4$ are known analytically
\cite{sphereanal,sphereanal2}.  It holds
\begin{eqnarray}
B_1&=&1\,,\label{39a}\\
B_2&=&4\,,\label{40}\\
B_3&=&10\,,\label{41}\\
B_4&=&\frac{2707\pi+438\sqrt{2}-4131\arccos(1/3)}{70\pi}=18.364768\dots\label{42}
\end{eqnarray}
The values are in the units of the packing fraction
 $\eta = \frac{\pi}{6}\sigma^3\rho$
with $\sigma$ standing for the sphere diameter.

The higher virial coefficients must be
calculated numerically.
Virial coefficients are sums of irreducible (cluster) integrals
represented by diagrams \cite{hill}. It is a task for a high
school student to determine the second virial coefficient for hard
spheres. However, dimensionality of irreducible integrals
increases rapidly with increasing order of the virial coefficient.
In three dimensional space it is $3n-6$. For example the tenth virial coefficient contains 24-dimensional integrals (!) in the
simplest case of spherically symmetric interactions between
molecules.

\begin{table}
\caption{Number of unlabeled and labeled Mayer and Ree-Hoover diagrams,
  and dimensionality $d$ of corresponding cluster integrals for hard
  spheres.}
\label{tab1}
\vspace{2ex}
\begin{center}
\begin{tabular}{|r|r|r|r|r|r|}\hline
$n$&  \multicolumn{2}{|c|}{Mayer}&\multicolumn{2}{|c|}{Ree-Hoover}&~~~$d$\\
\cline {2-5}
   & ~~~~unlabeled &  labeled & ~~~~unlabeled &  labeled&
\\
\hline\hline
 2  &        1&                1&       1&                 1& 1\\
 3  &        1&                1&       1&                 1& 3\\
 4  &        3&               10&       2&                 4& 6\\
 5  &       10&              238&       5&                68& 9\\
 6  &       56&          11\,368&      23&            3\,053&12\\
 7  &      468&      1\,014\,888&     171&          297\,171&15\\
 8  &   7\,123&    166\,537\,616&  2\,606&      56\,671\,216&18\\
 9  & 194\,066& 50 680\,432\,112& 81\,564& 21\,286\,987\,064&21\\\hline
\end{tabular}
\end{center}
\end{table}

\looseness=1 Numbers of integrals (no general formula is known for the number of
irreducible diagrams in dependence on $n$) increase even much more
rapidly than the dimensionality of the integrals, see
table~\ref{tab1}. Some of them are topologically equivalent which means
that their integrands differ only in the numbering of variables. In the
table, the heading ``labeled'' denotes the total number of irreducible
diagrams and ``unlabeled'' denotes the number of topologically different
diagrams with numbered black points. To reduce these numbers, Ree and
Hoover in their pioneering work \cite{sixth} replaced Mayer diagrams
with $f$-bonds by generalized Ree-Hoover diagrams with $f$-bonds and
$e$-bonds using the identity $f(r)+e(r)=1$. Besides the reduction of the
number of topologically different diagrams, there is another advantage
of the Ree-Hoover approach: computer codes using these diagrams are more efficient
than the Mayer approach.

For higher virial coefficients, say $n>7$, it is impossible within an
average lifetime of an explorer to determine the diagrams and their
weights using ``pencil and paper'' avoiding ``human factor'' errors. For
example, for $B_8$ there are 7\,123 different Mayer diagrams which can
be reduced to 2\,606 Ree-Hover diagrams, and for each of them their
weights should be determined.  The analysis should be done automatically on
a computer. Different algorithms utilizing symbolic algebra programming
and automatic code generation have been proposed
\cite{eight2,ninth,tenth,tenth2}.

Another problem is to evaluate the multi-fold integrals. This can
be done by random-shooting Monte Carlo integration but it may
require too much computer time. A more efficient way is to start
from the so-called \index{spanning diagram} \emph{spanning
diagrams} -- the integrals that can be calculated analytically.
Configurations of the spanning diagrams are sampled either using
a standard Metropolis Monte Carlo method or (for the linear
chains) by the so-called \emph{reptation} \cite{2tildesley}.  The
values of the diagrams of interest are then evaluated by Monte
Carlo integration.

\begin{table}[!h]
\caption{Summary of recommended virial coefficients $B_n$ for hard
  spheres. Values in parentheses are standard errors in the last
  significant digit.}
\vspace{2ex}
\label{tab2}
  \begin{center}
  \begin{tabular}{|c|c|}\hline
    \multicolumn{1}{|c|}{$n$}&
    \multicolumn{1}{|c|}{$B_n$}\\
    \hline\hline
   5  & 28.22445(10)\\
   6  & 39.81545(34) \\
   7  & 53.3418(15)  \\
   8  & 68.5394(87)   \\
   9  & 85.805(58)    \\
   10 &105.8(4)  \\
   \hline
  \end{tabular}
  \end{center}
 \end{table}
The fifth virial coefficient for hard spheres was calculated by
Rosenbluth and Rosenbluth \cite{fifth1} and by Kratky \cite{fifth2,fifth2_2,fifth2_3,fifth2_4}, the
sixth by Ree and Hoover \cite{sixth}, the seventh by Ree and Hoover \cite{seventh1}, by Kim and Henderson
\cite{seventh2}, and by Janse van Rensburg and Torrie
\cite{seventh3}, the eighth by Janse
van Rensburg \cite{eight1} and by Vlasov et al.\ \cite{eight2}, the
ninth by Lab\'\i k et
al.\ \cite{ninth}, and the tenth by Clisby and McCoy \cite{tenth,tenth2}. As a
rule, while higher virial coefficients were just calculated, the lower
ones were more precisely recalculated. The state-of-the-art values of
the virial coefficients for hard spheres are shown in table~\ref{tab2}
including their estimated uncertainties.

\section{Results and discussion}

\subsection{Extrapolation of Virial Coefficients}\label{etrapolovanevirialy}

It is prohibitively difficult to calculate virial coefficients beyond
$B_{10}$ because the numbers of diagrams increase enormously.  There are
$2^{n\choose 2}$ diagrams with $n$ field points that should be
analyzed. It is $2^{45}$ for $n=10$, $2^{55}$ for $n=11$, $2^{66}$ for
$n=12$, etc. The number of diagrams increases by more than three
orders with $n$ going from 9 to 10. It is thus practical to estimate
higher virial coefficients for $n>10$ rather than to try to calculate
them.

There are several approaches to extrapolation of virial coefficients based on
an (unjustified) assumption that the higher coefficients depend on the lower ones. In
the theory of fluids the most popular are Pad\'e approximants, see
 \cite{sanches} and references therein. They are based on the
assumption that the compressibility factor may be expressed in the form
\begin{equation}
  Z=\frac{P_n(\eta)}{Q_m(\eta)}\,,\label{46}
\end{equation}
where $P_n$ and $Q_m$ are polynomials of the order $n$ and $m$,
respectively, whose coefficients are combinations of the known virial
coefficients. By expanding equation~(\ref{46}) into the Taylor series,
higher virial coefficients are obtained.

Differential approximants
\cite{guttmann} are an extension of Pad\'e approximants. The first order differential approximants have the form
\begin{equation}
 P_n(\eta)\frac{{\rm d}Z}{{\rm  d}\eta}+Q_m(\eta)Z(\eta)=R_m(\eta)\,,\label{47}
\end{equation}
where $P_n$, $Q_m$, and $R_m$ are polynomials; for $P_n(\eta)=0$,
the Pad\'e approximants are revealed. Differential
approximants have not been used in the theories of fluids so far.

These two approaches utilize only a part of available information,
namely the lower virial coefficients. Another source of
information are computer simulation data on the compressibility
factors.  They can be utilized by proposing an equation of state
(EOS) some of its parameters being determined from the known virial
coefficients and the remaining constants fitted to the simulation
data. For example, Kolafa et al.\ \cite{jirkaEOS} used an equation
of state in the form proposed by Barboy and Gelbart \cite{barboy}
\index{equation of state (EOS)}
 \begin{equation}
 Z=1+\sum_{i=1}^n a_ix^i\,,\qquad x=\frac{\eta}{1-\eta}\,,\label{48}
 \end{equation}
 where lower $a_i$ are combinations of the known $B_i$ and the higher ones are fitted to the computer simulation data.
  By expanding the equation in powers of $\eta$,
 the higher virial coefficients may be estimated.

 This approach does not take into account the suggested
convergence limit at high densities \cite{kraska}. Therefore, we
extended the form (\ref{48}) by changing the pole $\eta=1$ into
\begin{equation}
  x = \frac\eta{\eta_{\rm pole}-\eta}\,,
  \label{pole}
\end{equation}
where $\eta_{\rm pole}$ is a (nonlinear) adjustable parameter.
This pronounced pole should not be confused with a weak
nonanalyticity at the freezing point discussed below.

All the traditional extrapolation methods suffer from an
assumption that
 the values of lower virial coefficients are known precisely. This
 is, certainly, not true, see table~\ref{tab2}. To be accurate,
 uncertainties in the lower virial coefficients (as well as in EOS simulation data)
 should be taken into account \cite{erpenbeck}.

 We believe that the safest way of utilizing both the
 lower virial coefficients data and the EOS data is to minimize the
 objective function
 \begin{equation}
 F=\sum_{i=5}^n \left(\frac{B_i-B_i^{\rm exp}}{\sigma_i}\right)^2+\sum_{j=1}^k
 \left(\frac{Z_j-Z_j^{\rm exp}}{\sigma_j}\right)^2\,,\label{49}
 \end{equation}
 where $n$ is a number of known virial coefficients, $\sigma_i$ are their
 standard deviations, $k$ is the number of simulated state points, and
 $\sigma_j$ are their standard deviations. $Z_j$ is the value of the
 compressibility factor at density $\eta_j$ given by a chosen compressibility factor
 in the form of equation~(\ref{46}) (Pad\'e approximants), or
 in the form of equation~(\ref{47}) (differential
 approximants), or in the form of equation~(\ref{48}) (Barboy-Gelbart equation).

As a measure of fit accuracy we use a standard deviation
$S$
\begin{equation}
S=\sqrt{\frac{F}{n-p}}\,,\label{50}
\end{equation}
where $p$ is a number of adjustable parameters.

\subsection{Barboy-Gelbart approach}

Let us return to equations (\ref{48}) and (\ref{pole}).  We allow for
zero coefficients $a_i$s in the expansions.  This improves precision
without adding more adjustable parameters especially at higher
densities.  For instance, equation (10) of \cite{jirkaEOS}, valid up to
$\rho=0.98$, uses powers $x^1$ to $x^8$ and $x^{12}$.  Therefore, the choice of
powers serves as another (discrete) adjustable parameter.

In order to elucidate the ability of expansions to predict unknown
virial coefficients we repeated the calculations with forms
(\ref{48}) and (\ref{49}), using various sets of input data.  That is, we
``forget'' that we know the virial coefficients up to $B_{10}$ and
use only a subset to $n$, $4\leqslant  n\leqslant 10$.

Moreover, we employ only subsets of available EOS data.  For
$\rho=0.94$ the error in $B_{11}\pm 1$ propagates to the error in $Z\pm
0.00083$ which is 8 times the standard deviation of the MD datum.  The
density range lower than $\rho_{\rm max}<0.94$, would not contain enough
information to allow for extrapolating the virial coefficients.  On the
other hand, the equation of state is nonanalytical at the freezing point
(density 0.947) \cite{jirka}, although the nonanalycity term is tiny and
in moderately metastable region it is not detectable within available
precision.  Therefore, we choose $\rho_{\rm max}=0.98$ as a compromise
and calculate all fits with five EOS sets from low densities up to
$\rho_{\rm max}=0.94, 0.95, 0.96, 0.97, 0.98$.
\begin{figure}[ht]
\begin{center}
  \includegraphics[scale=0.5]{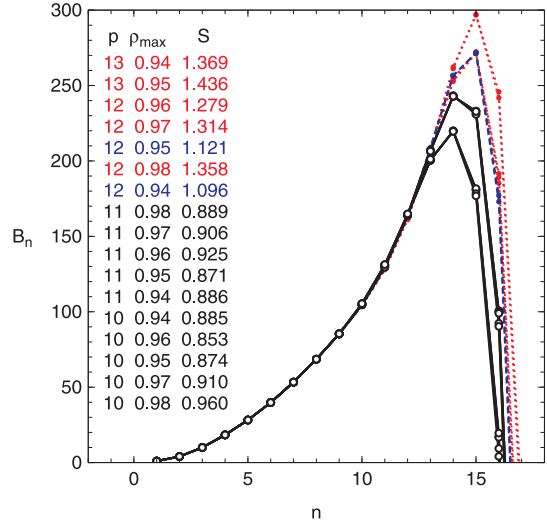}
\end{center}
\vspace{-1ex}
\caption{(Color online) Example of extrapolating the virial coefficients using formulas
  $Z=1+\sum_{i=1}^9 a_ix^i + a_px^p$ based on $B_2$ to $B_8$ and EOS
  data up to $\rho_{\rm max}$.  Virial coefficients $B_n$, $n>8$, are
  predicted, see the marked$^\dagger$ column of table~\ref{BGtab}.
  Black solid lines are the most precise fits ($S<1$), blue dashed lines
  are less precise ($S\in[1,1.2)$), and red dotted lines the least precise
  ($S\in[1.2,1.5)$). The table is ordered according to the predicted
  $B_{15}$ value and color-coded in the same way as the lines;
  some lines overlap.}
\label{B4+4R1}
\end{figure}

\begin{table}[!h]
\caption{Barboy-Gelbart predictions of $B_n$ using the equation of state
  up to reduced densities 0.94 to 0.98.  Knowledge of all virial
  coefficients up to $B_M$ was used, higher were predicted.  The error
  bounds in parentheses include both systematic and statistical
  contributions.}
\label{BGtab}
\vspace{2ex}
\begin{center}
\begin{tabular}{|c|c|c|c|c|c|c|c|c|}\hline
$n$ & $M=4$ & $M=5$ & $M=6$ & $M=7$ & $M=8^\dagger$ & $M=9$ & $M=10$\\\hline\hline
5   &28.255(30) &&&&&&\\
6   &39.67(7)   &39.716(23) &&&&&\\
7   &53.08(23)  &53.314(20) &53.31(12) &&&& \\
8   &69.0(2)    &69.36(9)   &68.6(2)    &68.66(5)   &&&\\
9   &87.7(10)   &87.6(5)    &85.8(5)    &85.74(14)  &85.30(8)   &&\\
10  &109(3)     &106.8(8)   &105.9(18)  &104.9(2)   &105.1(5)   &106.3(1)   &\\
11  &130(5)     &126.0(9)   &130.3(40)  &128(2)     &130.5(7)   &130.9(7)   &130.5(10)  \\
12  &150(3)     &142(2)     &160(5)     &158(7)     &164.2(9)   &156(3)     &156(3)     \\
13  &162(15)    &152(6)     &190(10)    &197(11)    &204(4) &174(9)     &176(4)     \\
14  &160(40)    &159(20)    &230(40)    &230(15)    &232(26)    &175(10)    &184(8)     \\
15  &130(70)    &161(30)    &180(100)   &240(20)    &210(40)    &165(50)    &179(32)\\  \hline
\end{tabular}
\end{center}
$^\dagger$ see figure~\ref{B4+4R1}
\end{table}
\begin{table}[!h]
\caption{The same as table~\ref{BGtab} with expansion (\ref{pole}).}
\label{poletab}
\begin{center}
\begin{tabular}{|c|c|c|c|c|c|c|c|c|}\hline
$n$ & $M=4$ & $M=5$ & $M=6$ & $M=7$ & $M=8^\dagger$ & $M=9$ & $M=10$\\\hline\hline
5   &28.240(35) &&&&&&\\
6   &39.73(11)  &39.72(2) &&&&&\\
7   &53.2(3)    &53.300(13)  &53.26(3) &&&&\\
8   &69.1(3)    &69.24(7)    &68.50(3)  &68.67(3) &&&\\
9   &87.3(14)   &87.4(4)     &86.0(3)   &85.66(8)    &85.48(9) &&\\
10  &107(4)     &107.0(7)    &106.4(7)  &104.80(16)  &105.3(3)    &106.15(6) &\\
11  &129(5)     &126.7(5)    &131.2(4)  &127.7(6)    &130.38(12)  &130.8(2)   &130.6(3)  \\
12  &148(9)     &144.7(8)    &162(3)    &157(4)      &162.2(24)   &158.1(14)  &158(2)    \\
13  &164(9)     &158(4)      &192(6)    &193(8)      &205(13)     &182(4)     &182(3)    \\
14  &174(35)    &165(12)     &207(16)   &227(11)     &234(35)     &192(7)     &192(4)    \\
15  &170(120)   &145(45)     &175(20)   &240(10)     &200(15)     &170(20)    &171(8)    \\\hline
\end{tabular}
\end{center}
$^\dagger$ see figure~\ref{B4+4R1}
\end{table}
The number of adjustable parameters is important in obtaining relevant
predictions.  We started with a small set and increased the number of
adjustable parameters until the value of $S$ dropped below
$1.5$. However, we never used more parameters if $S<1$ was already
reached to avoid a fitting noise.  For each set of virial coefficients up
to $M$, $4\leqslant  M\leqslant  10$, we analyzed several maximum powers in expansion
(\ref{48}) with all the above mentioned EOS data sets.  The results were
sorted into three categories: the best with $S<1$, the good ones with
$S<1.2$, and the worst results with $S<1.5$.  Then, we plotted the curves of
the predicted $B_n$ vs.\ $n$ and graphically determined the median value and
also tried to estimate the uncertainty from the data scattering.  This
error estimate is rather sensitive to the systematic errors which are
larger than the statistical ones.  We admit a subjective
factor present in this procedure.  One example is shown in figure~\ref{B4+4R1}.

\looseness=-1 The results, based on more than a thousand of  equations analyzed,
 are collected in tables~\ref{BGtab} and \ref{poletab}.
It is seen that the procedure is capable of predicting the virial coefficients
even though the errors are sometimes underestimated.  Of course, the
virial coefficients obtained are less accurate than the directly
obtained MC values. Only for $B_{10}$, the precision of the predicted
value based on all previous coefficients approaches the precision of the
MC result \cite{tenth,tenth2}.  In addition, the accuracy of $B_{14}$ and
especially $B_{15}$ decreases.

\subsection{Pad\'e approximants}

Simultaneous fitting of MD data and virial coefficients to
rational functions (Pad\'e approximants) is much more difficult
because the equations are highly nonlinear.  The approximants often
give oscillating solutions from $B_{12}$ or $B_{13}$; we omit
these solutions because we assume that the coefficients should
behave ``regularly''.

First we investigated the rational functions using all the known
information, i.e., the exact values of $B_2$ to $B_4$, approximate
values of $B_5$ to$B_{10}$ with standard deviations, and MD data to
$\rho_{\rm max}=0.94$, 0.95, 0.96, 0.97, 0.98 (five tested variants).
Approximants Pade [2,12], [8,5], [3,10], and [9,3] are the best
(interestingly, [9,3] requires by 2 less parameters than [2,12]; [2,11]
is worse, $S=1.4$, pushing $B_{14}$ to lower values) which with the five
maximum densities gives 20 curves from which the predictions and their
errors were determined.  To test the predictive power of the procedure,
we repeated the calculations without knowledge of $B_7$ to $B_{10}$, see
table~\ref{Padetab}.  It is seen that the Pad\'e approximants predict
worse values of the lower virial coefficients than the Barboy and
Gelbart approach. Nevertheless, the predictions of $B_{11}$ to $B_{15}$
using all the data are comparable.
\begin{table}[ht]
  \caption{Predictions using Pad\'e approximants. See table~\ref{BGtab}
  for the explanation of symbols.}
\label{Padetab}
\begin{center}
\begin{tabular}{|c|c|c|}\hline
$n$ & $M=6$     & $M=10$       \\\hline\hline
7   & 53.28(7)  &              \\
8   & 68.94(16) &              \\
9   & 84.7(14)  &              \\
10  & 104.0(13) &              \\
11  & 130.6(4)  & 130.5(10)    \\
12  & 162(15)   & 158(3)       \\
13  & 195(30)   & 181(4)       \\
14  & 200(20)   & 196(10)      \\
15  & 150(60)   & 198(24)      \\   \hline
\end{tabular}
\end{center}
\end{table}
\vspace{-5mm}

\subsection{Reliability of extrapolations}

While the statistical errors in the predicted higher virial
coefficients can be easily evaluated, this cannot be said about
the systematic errors introduced by a particular functional form.
The only way to assess the systematic errors is to compare as many
functional forms (Pad\'e indices, powers in polynomials, etc.) as
possible. Then, we subjectively estimate (after cancelling the evidently
defective values) the ``most probable'' values of
them.

\begin{table}[ht]
\caption{The extrapolated fourteenth virial coefficient using all the known
  virial coefficients and densities up to $\rho=0.94$.  BG+1 denotes the
  Barboy-Gelbart equation (\ref{48}) with one additional term $x^{10}$,
  BG+11 with $x^{10}$ and $x^{11}$, etc., and no optimization of the
  virial coefficients values, O- in front of symbols means that the full
  optimization, (\ref{49}), was performed. Pad\'e and O-Pad\'e denote
  the nonoptimized and optimized Pad\'e approximants, equation~(\ref{46}).
  O-DA is the optimized differential approximant.}
\label{B14tab}
\vspace{2ex}
\begin{center}
\begin{tabular}{|l|r|r|}\hline
Method&$B_{14}$&$S$\\
\hline\hline
BG+1        & $-270$ & 234 \\
BG+11       & 55   & 19 \\
BG+111      & 198  &  1.54 \\
O-BG+1      & 215  & 1.59 \\
O-BG+11     & 193  & 1.45 \\
O-BG+111    & 186  & 0.97 \\
Pade [5,4]  & 204  & 8.3 \\
Pade [2,7]  & 198  & 11.3 \\
O-Pade [5,4]&  207 & 6.9 \\
O-Pade [2,7]&  206 & 6.7 \\
O-Pade [9,1]&  204 & 6.5 \\
O-Pade [7,3]&  201 & 6.3 \\
O-DA    &200& 6.3 \\
\hline
\end{tabular}
\end{center}
\vspace{-5mm}
\end{table}

An example of a comparison of different methods of extrapolation is
shown in table~\ref{B14tab} for the fourteenth virial coefficient. In the
table, ten particular extrapolation results are shown: three utilizing
the Barboy-Gelbart equation (\ref{48}) (with and without adjustable
virial coefficients), Pad\'e approximants, and a differential approximant,
equation~(\ref{47}).

The final recommended values of higher virial coefficients
including the estimated uncertainties are given in table~\ref{tab5}. They are based on critical examination of several
methods and many equations of state. In the table, our values are
compared  with recent literature estimates obtained independently
using different approaches by different authors
\cite{tenth,tenth2,tian,Bannermann,hu,Molerlo}. Except $B_{15}$ they
agree with ours within estimated errors. However, a general trend
is that we predict $B_{11}$ to $B_{13}$ higher whereas $B_{14}$
and $B_{15}$ lower.

\begin{table}[ht]
\caption{Extrapolated virial coefficients.}
\label{tab5}
\begin{center}
\begin{tabular}{|l|l|l|l|l|l|l|}\hline
 $n$ &this work&ref.~\cite{tenth,tenth2}&ref.~\cite{tian}&ref.~\cite{Bannermann}&ref.~\cite{hu}&ref.~\cite{Molerlo}\\
  \hline\hline
11 &\textbf{130.6(10)}  &128    &128    &129    &126    &128\\
12 &\textbf{157(3)} &153    &153    &154    &149    &153\\
13 &\textbf{185(10)}    &181    &181    &182    &174    &182\\
14 &\textbf{190(20)}    &215    &213    &211    &202    &214\\
15 &\textbf{180(30)}    &247    &248    &236    &231    &247\\
16 & --     &279    &288    &254    &262    &279\\
 \hline
\end{tabular}
\end{center}
\end{table}
In the table our estimated values of $B_n$ are shown up to $n=15$.
We calculated higher virial coefficients as well. However, their
values exhibited a large scatter and no final estimates were thus
made. Anyway, the literature estimates for $B_{16}$ are given in
the table.

We cannot assert that really $B_{15}<B_{14}$; in fact, oscillating
behavior is typical of polynomial approximants and these results
are consistent with a monotonous increase, at least for not too
large $n$. Ultimately the sequence of $B_n$ cannot be monotonous
because the equation of state is not analytical at the freezing
point and consequently its radius of convergence cannot exceed the
freezing density (we believe that it equals the freezing density).

\section{Concluding remarks}

In this work we have used several extrapolation methods to estimate the
unknown virial coefficients of hard spheres from $B_{11}$ to $B_{15}$
and compared them with recent literature estimates. We believe that the
presented results could be used in testing theoretical approaches of the
thermodynamics of hard spheres.

The problem of reliability of any extrapolation result rests in the
estimate of its uncertainty.  This is, in principle, subjective
depending on an author's view (either optimistic or pessimistic).  Here
we have tried neither to be too optimistic nor too pessimistic.  We have
used comparisons of extrapolated virial coefficients obtained using as
many different approaches as possible.

The techniques used in this work can be readily extended to virial
coefficients of hard body fluids such as spherocylinders, diatomics and
multi-atomics, mixtures of hard spheres and other hard body fluids and
their mixtures. What one needs is i) reliable values of lower virial
coefficients and estimates of their uncertainties, and ii) precise
equation-of-state computer simulation data. An extrapolation of virial
coefficients of non-hard-body fluids is less straightforward due to
their dependence on temperature.

\ukrainianpart

\title{До екстраполяції віріальних коефіцієнтів твердих сфер}
\author{М. Ончак\refaddr{A1}, А. Малієвскі\refaddr{A1,A2}, Й. Колафа\refaddr{A1}, С. Лабік\refaddr{A1}}

\addresses{
\addr{A1}{Інститут хімічної технології, Прага, Відділ фізичної
хімії, 166~28 Прага~6, Чеська Республіка}
\addr{A2}{Відділ фізики,
факультет природничих наук, університет м. Острава, 701 03 Острава
1, Чеська Республіка}
}

\makeukrtitle

\begin{abstract}
\tolerance=3000%
Обговорюються декілька екстаполяційних методів для віріальних
коефіцієнтів, включно із запропонованими в цій роботі. Показано
можливості методів передбачати вищі віріальні коефіцієнти
однокомпонентної системи твердих сфер. Запропоновано оцінки для
віріальних коефіцієнтів від одинадцятого до п'ятнадцятого.
Передбачається, що віріальні коефіцієнти $B_n$ вищого порядку ніж
$B_{14}$, можуть зменшуватися з ростом $n$ і можуть досягати
негативних значень для великих $n$. Ці екстраполяційні методи можуть
бути використані в інших областях, де екстраполяція є важливою.

\keywords тверді сфери, віріальні коефіцієнти, екстраполяційні
методи

\end{abstract}


\newpage
\begin{thebibliography}{99}
\bibitem{hill}
  Hill~T.L., Statistical Mechanics, McGraw-Hill, New York, 1956.
\bibitem{sphereanal}
Nijboer~B.R.A., Van Hove~L., Phys. Rev., 1952, \textbf{85}, 777; \doi{10.1103/PhysRev.85.777}
\bibitem{sphereanal2}
Lyberg I., e-print cond-mat/0410080.

\bibitem{sixth}
Ree~F.H., Hoover~W.G., J. Chem. Phys., 1964, \textbf{40}, 939; \doi{10.1063/1.1725286}.

\bibitem{eight2}
Vlasov~A.Y., You X.-M., Masters A.J., Mol. Phys., 2002, \textbf{100}, 3313; \doi{10.1080/00268970210153754}.

\bibitem{ninth}
Lab\'\i k S., Kolafa J., Malijevsk\'y A., Phys. Rev. E, 2005, \textbf{71}, 021105; \doi{10.1103/PhysRevE.71.021105}.

\bibitem{tenth}
Clisby N., McCoy~B.M., J.~Stat. Phys., 2006, \textbf{122}, 15; \doi{10.1007/s10955-005-8080-0}
%
\bibitem{tenth2}
Clisby N., McCoy~B.M., Pramana-J. Phys., 2005, \textbf{64}, 775; \doi{10.1007/BF02704582}.

\bibitem{2tildesley}
Allen M.P., Tildesley~D.J., Computer Simulation of Liquids.
Clarendon Press, Oxford, 1987.

\bibitem{fifth1}
 Rosenbluth~M.N., Rosenbluth~A.W., J.~Chem. Phys., 1955, \textbf{23}, 356; \doi{10.1063/1.1741967}.

\bibitem{fifth2}
Kratky~K.W., Physica A, 1976, \textbf{85}, 607; \doi{10.1016/0378-4371(76)90029-7}.

%
\bibitem{fifth2_2}
Kratky~K.W., Physica A, 1977, \textbf{87}, 584; \doi{10.1016/0378-4371(77)90051-6}.

%
\bibitem{fifth2_3}
Kratky~K.W., J.~Stat. Phys., 1982, \textbf{27}, 533; \doi{10.1007/BF01011091}.
%
\bibitem{fifth2_4}
Kratky~K.W., J.~Stat. Phys.,  1982 \textbf{29}, 129; \doi{10.1007/BF01008253}.

\bibitem{seventh1}
Ree~F.H., Hoover~W.G., J. Chem. Phys., 1967, \textbf{46}, 4181; \doi{10.1063/1.1840521}.

\bibitem{seventh2}
Kim S., Henderson D., Phys. Lett.~A, 1968, \textbf{27}, 378; \doi{10.1016/0375-9601(68)91066-9}.

\bibitem{seventh3}
van Rensburg~E.J., Torrie~G.M., J. Phys. A: Math. Gen., 1993, \textbf{26}, 943; \doi{10.1088/0305-4470/26/4/022}.

\bibitem{eight1}
van Rensburg~E.J., J. Phys. A: Math. Gen., 1993, \textbf{26}, 4805; \doi{10.1088/0305-4470/26/19/014}.

\bibitem{sanches}
Sanchez~I.C., J.~Chem. Phys., 1994, \textbf{101}, 7003; \doi{10.1063/1.468456}.

\bibitem{guttmann}
 Guttmann~A.J. -- In: Phase Transitions and Critical
 Phenomena, Vol. 3, chapter 1, eds. C. Domb and J. Lebowitz.
 Academic Press, 1989.

\bibitem{jirkaEOS}
Kolafa J., Lab\'\i k S., Malijevsk\'y A., Phys. Chem. Chem. Phys.,
2004, \textbf{6}, 2335; \doi{10.1039/b402792b}.

\bibitem{barboy}
Barboy B., Gelbart~W.M., J.~Chem. Phys., 1979, \textbf{71}, 3053; \doi{10.1063/1.438711}.

\bibitem{kraska}
Yelash~L.V., Kraska T., Phys. Chem. Chem. Phys.,  2001, \textbf{3}, 3114; \doi{10.1039/b102972j}.

\bibitem{erpenbeck}
Erpenbeck~J.J., Wood~W.W., J.~Stat. Phys., 1984, \textbf{35}, 321; \doi{10.1007/BF01014387}.

 \bibitem{jirka}
Kolafa J., Phys. Chem. Chem. Phys., 2006, {\textbf 8}, 464; \doi{10.1039/b511999e}.

 \bibitem{tian}
 Tian~J.X., Gui~Y.X., Mulerlo A., Phys. Chem. Chem. Phys., 2009, \textbf{11}, 11213; \doi{10.1039/b915002a}.

 \bibitem{Bannermann}
 Bannerman~M.N., Lue L., Woodcock~L.V., J. Chem. Phys., 2010, \textbf{132}, 084507; \doi{10.1063/1.3328823}.

 \bibitem{hu}
 Hu~J.W., Yu~Y.X., Phys. Chem. Chem. Phys., 2009,  \textbf{11}, 9382; \doi{10.1039/b911901a}.

 \bibitem{Molerlo}
 Tian~J.X., Gui~Y.X., Mulerlo A., J. Phys. Chem. B, 2010,  \textbf{114}, 13399; \doi{10.1021/jp106502x}.

\end{thebibliography}
\end{document}